\documentclass[%
 reprint,
 amsmath,amssymb,
 aps,
 prl,
]{revtex4-1}

\usepackage{graphicx}
\usepackage{dcolumn}
\usepackage{bm}

\graphicspath{{./img/}}

\usepackage{xcolor}

\newcommand{\beq}{\begin{equation}}
\newcommand{\eeq}{\end{equation}}
\newcommand{\beqa}{\begin{eqnarray}}
\newcommand{\eeqa}{\end{eqnarray}}
\newcommand{\bfig}{\begin{figure}\begin{center}}
\newcommand{\efig}{\end{center}\end{figure}}
\newcommand{\btab}{\begin{table}\begin{center}}
\newcommand{\etab}{\end{center}\end{table}}

\newcommand{\eqn}[1]{\mbox{Eq.\hspace{1pt}(\ref{#1})}}
\newcommand{\eqs}[2]{\mbox{Eq.\hspace{1pt}(\ref{#1}--\ref{#2})}}

\newcommand{\eqtn}[2]{\begin{equation} \label{#1} #2 \end{equation}}

\def\brp{{\mathbf{r}^{\prime}}}

\def\br{{\mathbf{r}}}

\def\d{{\mathrm{d}}}

\def\rhoir{{\rho_I({\bf r})}}

\def\sumi{{\sum_I^{N_S}}}

\def\sumij{{\sum_{J\neq I}^{N_S}}}

\begin{document}


\title{Cooperation and Environment Characterize the Low-Lying Optical Spectrum of Liquid Water}

\author{Sudheer Kumar P.}
\author{Michele Pavanello}
 \email{m.pavanello@rutgers.edu}
\affiliation{%
Department of Chemistry, Rutgers University, Newark, NJ 07102
}%

\date{\today}

\begin{abstract}
The optical spectrum of liquid water is analyzed by subsystem time-dependent density functional theory. We provide simple explanations for several important (and so far elusive) features. Due to the disordered environment surrounding each water molecule, the joint density of states of the liquid is much broader than that of the vapor. This results in a red shifted Urbach tail. Confinement effects provided by the first solvation shell are responsible for the blue shift of the first absorption peak compared to the vapor. In addition, we also characterize many-body excitonic effects. These dramatically affect the spectral weights at low frequencies, contributing to the refractive index by a small but significant amount.
\end{abstract}

\pacs{31.15.E−, 33.80.−b, 71.15.−m, 82.80.Pv}
\keywords{DFT, embedding, liquid water, optical spectrum}
\maketitle

Water is the most important liquid on Earth. Thus, understanding its optical spectrum is of pivotal importance. Although there has been tremendous progress in this area, difficulties arise in the ab-initio models of the liquid because large simulation cells need to be employed due to its disordered nature. In turn, this either forces the use of approximate methods on large cells, or accurate methods on cells that are too small. As a result of this limitation, there are several open questions and interesting features of the optical absorption spectrum of water that are yet to be fully explained. In this work, we explore two themes: (1) Many-body excitonic interactions between the water molecules, and (2) coupling of the first absorption band to the nuclear degrees of freedom describing the liquid structure.

By many-body effects, we mean the effects that arise when single water molecules (single bodies) in the liquid interact with each other (other bodies) both in the ground state as well as in their excited states. Although related, this definition is different in spirit from the kinds of many-body effects that a Bethe-Salpeter Equation (BSE) treatment would recover. In the latter, many-body refers to electron--hole interactions. We set out to investigate how many-body interactions affect the optical spectrum and other related quantities (such as the refractive index). These can be cooperative or anticooperative in nature. 

Many-body effects have been discussed before in terms of the screening properties of the bulk in the computation of self-energies for GW calculations \cite{Garbuio_2006,Ziaei_2016}. It was found that screening is independent of the particular configuration of water considered. Thus, it can be inferred that it is not affected by the structure of the environment surrounding the water molecules. It has also been shown that for ice, cooperative many-body effects in the form of excitonic couplings increase the oscillator strength of low-lying excitations and are responsible for an increase of the index of refraction with increasing pressure \cite{Pan_2014}.

In addition, we also set out to investigate coupling between the first absorption band and the structure of the liquid. This helps us understand what influences the peak position and shape (broadening). The underlying reasons for a blue shift of the first absorption band of water from 7.4 eV in gas phase \cite{Chan_1993} to 8.3--8.6 eV in the liquid phase \cite{Elles_2009,Heller_1974,Hayashi_2015} are still matter of debate. G$_0$W$_0$ calculations based on PBE orbitals predict a shift of about 0.7 eV and attribute it to excitonic effects \cite{Ziaei_2016,Hahn_2005} (i.e., the isolated water molecule features a more localized exciton than that of the liquid phase). Calculations based on clusters, instead, define the blue shift as a result of electrostatic embedding \cite{Hermann_2008} (a sort of confinement effect). This view is corroborated by excitonically coupled coupled cluster (CCSD) and semilocal time-dependent Density Functional Theory (TDDFT) calculations showing that the first absorption band is composed of localized states \cite{Mata_2009,Lu_2008}. A recent analysis of the non self--consistency of the previously mentioned G$_0$W$_0$ results uncovered some possible biases that are carried over from using DFT orbitals in the GW+BSE part of the calculation \cite{Blase_2016} putting into question the previously predicted exciton sizes of liquid and vapor.
\begin{figure}
\includegraphics[width=0.5\textwidth]{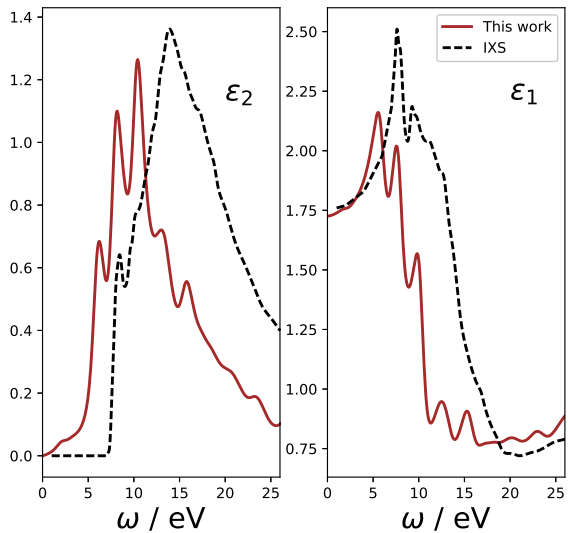}
\caption{\label{fig1} Real and imaginary part of the frequency dependent dielectric constant of liquid water. IXS stands for inelastic X-ray scattering data from Ref.\citenum{Hayashi_2015}. In red our subsystem TDDFT results, which were computed with a large broadening parameter ($\eta=0.5$ eV) to better compare to the experiment.}
\end{figure}
%
%

We set out to attack these open questions with simulations based on subsystem DFT \cite{jaco2014,weso2015,krish2015a}. 
The electron density is decomposed into subsystem contributions, $\rho(\br) = \sumi \rhoir$, where $N_S$ is the number of subsystems. The subsystem densities are recovered variationally solving subsystem-coupled Kohn--Sham equations, $\left( -\frac{1}{2}\nabla^2 + v_s^I(\br) + v_{\rm emb}^I(\br) \right)\phi_i^I(\br) = \varepsilon_i^I  \phi_i^I(\br)$, where $v_s^I$ is the Kohn--Sham potential of the isolated subsystem evaluated at $\rhoir$. The embedding potential, $v^I_{\rm emb}$, contains exact Coulomb interactions with surrounding subsystems, as well as functional derivatives of nonadditive exchange--correlation (xc) and nonadditive noninteracting kinetic energy functional ($T_s$) \cite{jaco2014,weso2015,krish2015a}. To access information about excited states, subsystem TDDFT can be formulated either in frequency \cite{casi2004} or time domain \cite{krish2015a}.
\begin{figure*}
\includegraphics[width=0.9\textwidth]{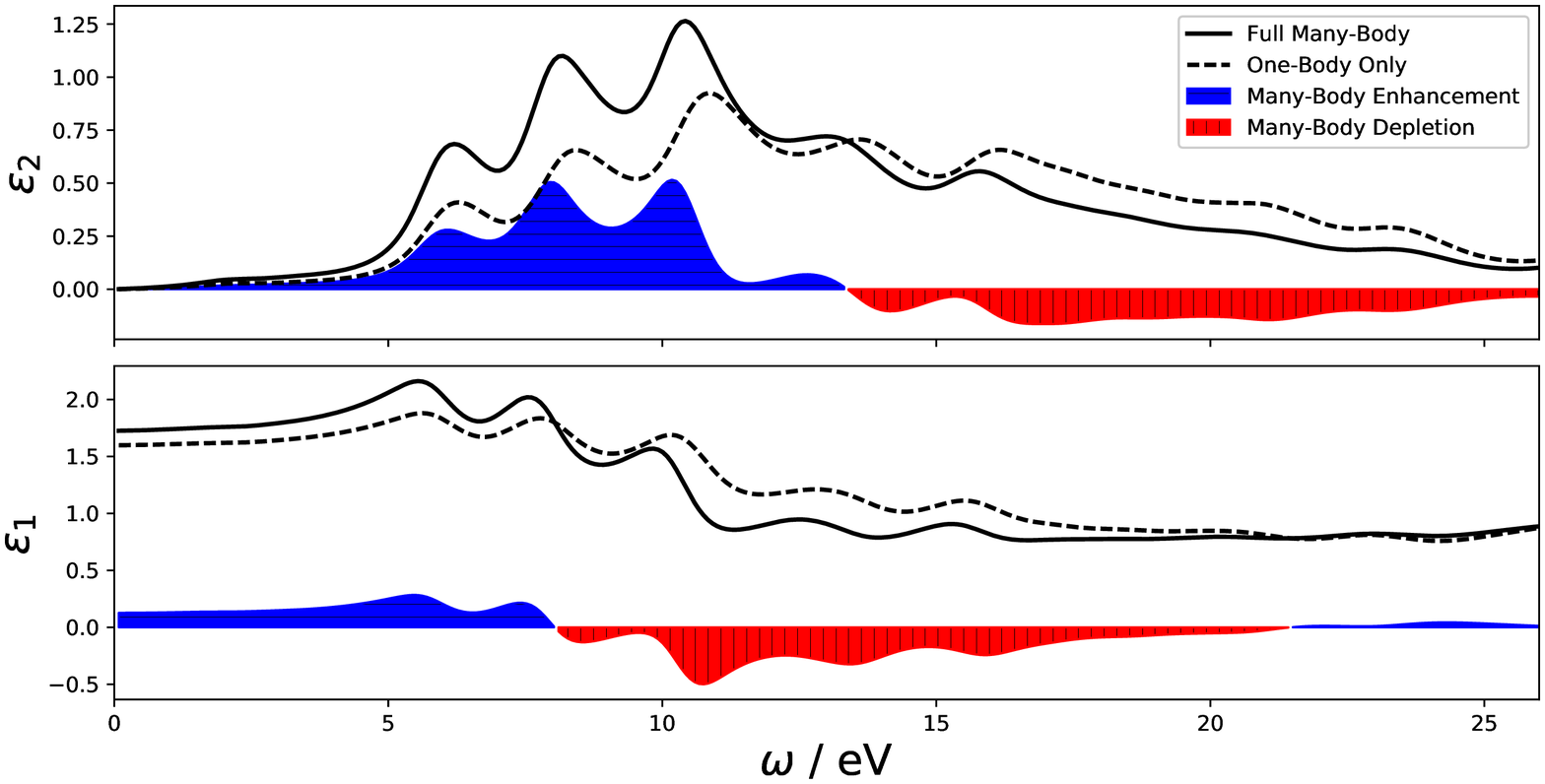}
\caption{\label{fig2} One-body (i.e., employing only the uncoupled response functions) and full many-body (i.e., employing \eqs{chi1}{chi2}) real and imaginary part of the frequency dependent dielectric constant of liquid water. Filled curves highlight the difference between the one-body and the full many-body result.}
\end{figure*}
Subsystem TDDFT allows us to approach larger supercells than before and naturally gives us access to many-body effects without compromising the accuracy of the model. Many-body effects arise as inter-subsystem static and time-dependent interactions \cite{krish2015a,neug2010a,krish2015b}. These perks are further complemented by the ability of subsystem DFT to naturally localize the electronic structure of the subsystems while still allowing electron density overlap. We have showed, for example, that it can quantitatively reproduce the dynamics and structure of liquid water because the too strong hydrogen bond resulting from self-interaction in semilocal xc functionals, is offset by the subsystem-local electronic structure \cite{Genova_2016a}.
While we do not expect that subsystem electron density localization be a generally beneficial feature of the model, it is beneficial in the limit of simulating systems composed of noncovalently bound subunits. Such as molecular liquids, crystals, and layered periodic systems \cite{eQE}.


Simulations were carried out with the package embedded Quantum-ESPRESSO (eQE) \cite{eQE}, employing the PBE xc functional and ultrasoft pseudopotentials (O.pbe-rrkjus.UPF/H.pbe-rrkjus.UPF from the Quantum-ESPRESSO PP Library \cite{Rappe_1990}). We employ plane wave kinetic energy cutoffs of 50 and 500 Ry for the waves and charge densities, respectively. Real-time subsystem TDDFT was implemented in the linear response regime, applying a $\delta$ electric field perturbation \cite{Yabana_1996} of 0.02 Ry/\AA , a time step of 1 as, and assuming the adiabatic approximation for all the density-dependent potentials (xc as well as nonadditive $T_s$). The time-dependent KS orbitals were propagated for 50,000 steps with a Crank-Nicholson propagator, totalling 50 fs of electron dynamics carried out for each of the three Cartesian directions. We considered a total of 8 snapshots of a subsystem DFT based Born-Oppenheimer dynamics of bulk liquid water modelled by 64 independent water molecules in a cubic box ($a=12.4278$\AA, presented in a previous work \cite{Genova_2016a}). We employ the LC94 \cite{lemb1994} nonadditive $T_s$ functional, which was shown to satisfactorily reproduce energy surfaces of CCSD(T) benchmarks for water dimer as well as the structural parameters of the liquid (e.g., radial and angular distribution functions) \cite{Genova_2016a}.

The real-time subsystem dipole change, $\delta \mu_I(t) = \int \br \left( \rho(\br,t) - \rho(\br,t=0) \right) \d\br$, is Fourier transformed to frequency space to yield the isotropic dipole polarizability, $\alpha_I(\omega)$ which is related to the frequency-dependent dielectric constant \cite{Hayashi_2015,Bertsch_2000}, 
$\epsilon(\omega) = \left[ 1+ 4\pi \langle\alpha(\omega)\rangle \right]^{-1}$, where the average is carried out over all water molecules for all snapshots, $\langle\alpha(\omega)\rangle = \langle \frac{1}{N_S} \sumi \alpha_I(\omega)\rangle_\text{Snapshots}$. The Fourier transformation was carried out including an artificial peak broadening (see supplementary information), and the sum rule for the polarizability was normalized to the experimental value in the range 0-25 eV. 

In Figure \ref{fig1}, we show the comparison between our calculation and the most recent experiment for the real ($\epsilon_1$) and imaginary ($\epsilon_2$) parts of the frequency-dependent dielectric constant. Although differences are evident, our subsystem TDDFT calculation recovers the overall trend and also reproduces some interesting features. The computed $\epsilon_2$ is consistently red shifted, however the overall multimodal shape is reproduced. For $\epsilon_1$, we satisfactorily reproduce the $\omega=0$ limit, as well as the peak maximum.

Subsystem TDDFT allows us to inspect the dynamical interactions between subsystems, and allows us to glimpse into the cooperative interactions that arise when the liquid responds to an external perturbation. A formally exact decomposition of the interacting electronic response function of a system into subsystem contribution \cite{casi2004,pava2013b,neug2007} can be implemented. Namely,
\eqtn{chi1}{\chi = \sumi \chi_I^c,}
with each interacting subsystem response function given by a local interacting one-body term ($\chi_I^u$) and a nonlocal many-body term,
\eqtn{chi2}{\chi_I^c = \underbrace{\chi_I^u}_\text{one body} + \underbrace{\sumij \chi_I^u K_{IJ} \chi_J^c,}_\text{many body}}
where hereafter superscripts $c$ and $u$ stand for ``coupled'' (i.e., employing the full many-body response) and ``uncoupled'' (i.e., employing only the one-body response).

The subsystem TDDFT kernel for $I\neq J$ is given by $K_{IJ}(\br_1,\br_2,t-t^\prime) = \frac{\delta(t-t^\prime)}{|\br_1-\br_2|} + \frac{\delta^2 E_{\rm xc}}{\delta\rho(\br_1,t)\delta\rho(\br_2,t^\prime)}  + \frac{\delta^2 T_{\rm s}}{\delta\rho(\br_1,t)\delta\rho(\br_2,t^\prime)}$ \cite{casi2004,neug2010a}. The dipole polarizability is related to the response function by $\alpha_{xy}(\omega)=-\int x^\prime \chi(\brp,\br,\omega)y \d\br\d\brp$. The uncoupled subsystem-local response function is computed with response equations that include occupied-virtual orbital excitations of only one subsystem. Or alternatively, in the real-time approach the embedding potential is computed at each time step only updating the subsystem time-dependent density and leaving the density of the surrounding subsystems frozen at $t=0$. The many-body terms include couplings between subsystem excitations at the level of two and higher bodies \cite{krish_2016}, $\chi_I^u K_{IJ} \chi_J^u +\chi_I^u K_{IJ} \chi_J^uK_{JK}\chi_K^u \ldots$

It is evident from Figure \ref{fig2} that the many-body contributions are cooperative at low frequencies, and anticooperative at high frequencies. Although it is difficult to pinpoint the specific reasons for this behavior, we note that increased spectral weights at low frequencies are typical of excitonically coupled systems \cite{Onida_2002}. We should remark that the many-body term in \eqn{chi2} must be associated with an overall zero sum rule \cite{krish_2016}. Thus, if there is a many-body enhancement in one region of the spectrum there must be a many-body depletion in another region of the spectrum. In addition, the predicted many-body enhancement at low frequencies is consistent with the finding that the refractive index in water increases with increasing pressure, ascribed to an increase of the oscillator strengths with pressure \cite{Pan_2014}. From our results, we estimate the refractive index as $n=\sqrt{\epsilon_1(\omega=0)}$, and obtain $n=1.25$ and $1.30$ for the one-body and the full many-body results. Thus, many-body effects increase $n$ by about 4\%. 
\begin{figure*}
\includegraphics[width=0.8\textwidth]{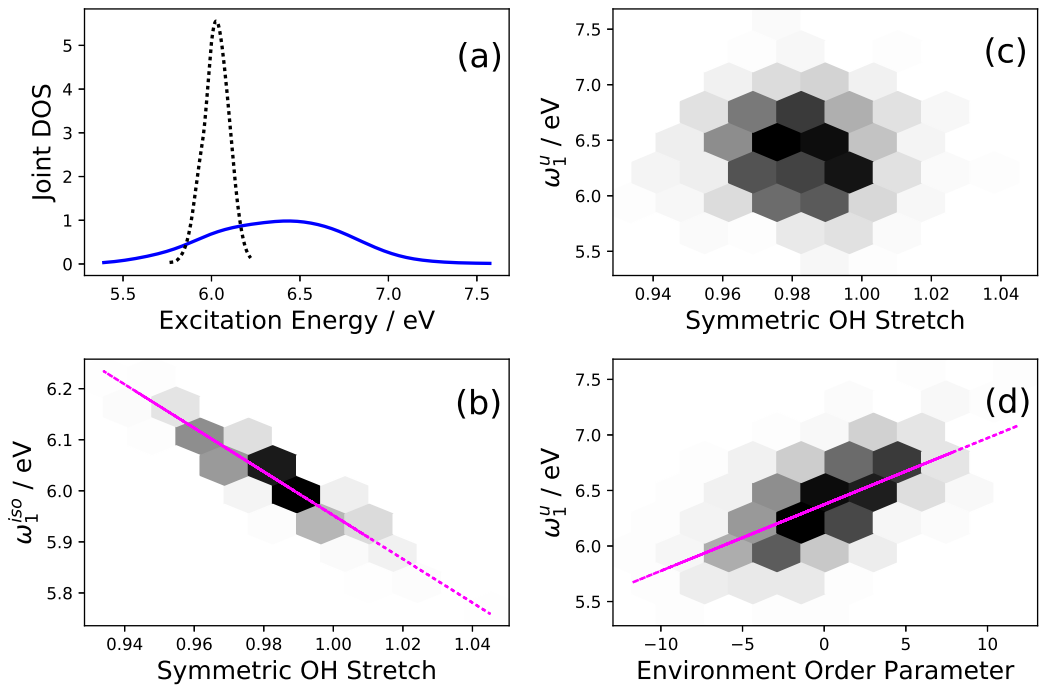}
\caption{\label{fig3} (a) Joint density of states of the first excitation energy of gas phase (dotted line) and first absorption band of liquid water (blue solid line); (b) First excitation energy of isolated water molecules  against the symmetric stretch mode; (c) First absorption band of liquid water molecules against the symmetric stretch mode; (d) First absorption band of liquid water molecules against the environmental order parameter. Linear fits of the correlation scatter plots (magenta lines) are shown. Insets (b-d) are represented by hexagonal bins. Darker hexagons correspond to higher count value.}
\end{figure*}

After having characterized the role of many-body dynamical interactions between water molecules in the liquid, we now analyze the first absorption band. Subsystem TDDFT places the average peak position of the first absorption band at $\omega_1^u=6.4$ eV. Compared to the experimental 8.3--8.6 eV, we underestimate it by $\sim$2 eV. This is expected \cite{Garbuio_2006}, as we employ the PBE semilocal xc functional and the adiabatic approximation. Due to the more localized electronic structure compared to semilocal TDDFT of the supersystem, it is also expected that subsystem TDDFT yields a larger gap (semilocal TDDFT of the supersystem finds a gap of $\sim5$ eV \cite{Tavernelli_2006}). For the same reasons, the subsystem TDDFT value for $\epsilon_1(\omega=0)=1.68$, slightly underestimates the DFT-PBE value of 1.72 \cite{Lu_2008}.

We find that the many-body terms have a negligible effect on the peak position of the first absorption band, which only shifts by about 0.02 eV when they are included. In other words, $\omega_1^c-\omega_1^u=-0.02$ eV. This points to a unique feature of liquid water, that is the many-body terms affect the magnitude of the dielectric constant but do not change the position of the first peak in the optical spectrum.

In Figure \ref{fig3}, we summarize our results pertaining to the first absorption band.  Inset (a) of the figure displays histograms of the first excitation energy on the $x$-axis and the subsystem count on the $y$-axis, for the liquid (in blue) and for the vapor (dotted line). These histograms can be interpreted as vibrationally modulated joint density of states (JDOS). The gas-phase data  is generated with water molecules having the same geometry as the liquid, but their electronic structure is computed in the absence of environment with the ADF \cite{teve2001a} software employing the PBE xc functional and a quadruple-$\zeta$ Slater-Type Orbital basis set. From the JDOS, we evince that the gas-phase first excitation energy is sharply centered around the average (there are 512 data points in the histogram) at $\omega^{iso}_1=6.0$ eV ($\sigma^{iso}_1=0.07$ eV, onset at 5.8 eV), while the subsystem TDDFT places the band maximum at $\omega^u_1=6.4$ eV featuring a very broad range of excitation energies ($\sigma^u_1=0.4$ eV, onset at 5.4 eV). Such a broadening is entirely due to interactions of the water molecules with their environment. Because the environment is heterogenous and dynamic in the liquid, a broad distribution of excitation energies emerges. Additionally, Figure 3a shows that the onset of the absorption band (Urbach tail) is {red shifted} compared to the isolated water molecule case by 0.4 eV, again due to vibronic coupling to environmental degrees of freedom. 

To justify the above claim, we provide in Figure 3b and 3c, the scatter plot of the computed excitation energy for each of the isolated and embedded (liquid phase) water molecules against the symmetric stretching degree of freedom. We find that the excitation energy for the isolated water molecules almost perfectly anticorrelates to the stretching mode. Conversely, the first excitation of the embedded water molecules does not correlate at all with this internal degree of freedom. Inspection of Figure 3b and 3c unequivocally determines that the symmetric stretching, which was determinant for the spectrum of gas phase water, no longer plays a role in the spectrum of the liquid. The other local degrees of freedom (asymmetric stretch and the bond angle) do not correlate to the absorption peak of either gas or liquid phases. 

 Thus, we set out to find the degree of freedom describing the environment surrounding a water molecule in the liquid that correlates the most to the first absorption peak of each subsystem. Among the ubiquitous order parameters (such as number of donated/accepted hydrogen bonds, coordination number, and tetrahedral order parameter \cite{DiStasio_2014b}), we found that only the number of accepted hydrogen bonds features a nonzero correlation to the first peak position. However, the computed correlation value ($\text{correlation}=\frac{\sigma(X,Y)}{\sqrt{\sigma(X,X)\sigma(Y,Y)}}$, with $\sigma$ being the variance and $X=\text{excitation energy}$, $Y=\text{degree of freedom}$) is only 0.3, and thus it is not large enough to confidently attribute the presence of a correlation. In Figure 3d we show the scatter plot of the first excitation energy of each subsystem against a composite order parameter which we call ``Environment Order Parameter'' (EOP) that includes degrees of freedom of molecules in the environment. EOP was constructed from 33 independent descriptors (O--O distances with the closest 6 waters and 12 respective angles with the hydrogen atoms, the O$_1$--O$_2$--H$_2$ and O$_2$--O$_1$--H$_1$  angles, $\theta$, the tetrahedral order parameter \cite{DiStasio_2014b}, accepted and donated number of hydrogen bonds). These initial descriptors were combined in nonlinear ways (for example, slicing the O--O distance in this way: $\text{O--O distance} \times (\theta-\theta_0)$, with $\theta_0$ being a constant angle value) to generate 268 linearly dependent environmental degrees of freedom. These were reduced by singular value decomposition of their covariance matrix (i.e., a principal component analysis) to only 29. This is a massive reduction in dimensionality, especially in view of the fact that the total number of possible binary nonlinear combinations of order parameters with 3 slices per set (i.e., 3 values of $\theta_0$) with 63 molecules in the environment equals to $63\times3\times62\times3=35154$ parameters. The 29 linearly independent order parameters resulting from the principal component analysis where linearly combined to yield EOP in such a way to maximize the correlation to the first excitation energy of each subsystem. The maximization was carried out with a multivariable optimization based on the BFGS algorithm. Correlation of the first absorption band peak position to EOP is 0.6. This is double the original best value from simple order parameters.

Analysis of the major components of EOP highlights four order parameters. In order of contribution: the distances to the 4-th (negative correlation), 6-th (negative correlation), 1-st (positive correlation) and 2-nd (negative correlation) oxygen atoms in the environment, all multiplied by $\theta-60^o$ (i.e., the O$_1$--O$_2$--H$_2$ angle). This explains the not large enough correlation to the number of hydrogen bonds, because the hydrogen bond definition is too simplistic and only combines a distance cut-off criterion in conjunction with an angle cut-off therefore missing the complexity of the interaction. In the supplementary information, we provide further details about the procedure to obtain EOP.

In conclusion, we carried out real-time subsystem TDDFT simulations of liquid water and compared the results to the gas-phase water. We show that, although many-body excitonic effects have no effect on the position of the first absorption band peak, they dramatically enhance the spectral intensities in the low frequency range, and deplete the intensities at high frequencies. The simulations also predict the liquid's Urbach tail to be 0.4 eV red shifted compared to the gas phase, due to the strong coupling of the first absorption band with the environment within the first solvation shell. Interaction with the environment thus dramatically broadens the joint density of states and blue shifts the band peak by 0.4 eV. The results reproduce semiquantitatively the experimental red shift of the Urbach tail and the blue shift of the band peak.

We thank Robert A. DiStasio, Jr. for helpful discussions, particularly for pointing out the need to slice the O--O distances to enhance correlations. We also thank Alessandro Genova for helping us with the dipole Fourier transform script. This material is based upon work supported by the U.S. Department of Energy, Office of Basic Energy Sciences, under Award Number [award number].

\begin{thebibliography}{31}%
\makeatletter
\providecommand \@ifxundefined [1]{%
 \@ifx{#1\undefined}
}%
\providecommand \@ifnum [1]{%
 \ifnum #1\expandafter \@firstoftwo
 \else \expandafter \@secondoftwo
 \fi
}%
\providecommand \@ifx [1]{%
 \ifx #1\expandafter \@firstoftwo
 \else \expandafter \@secondoftwo
 \fi
}%
\providecommand \natexlab [1]{#1}%
\providecommand \enquote  [1]{``#1''}%
\providecommand \bibnamefont  [1]{#1}%
\providecommand \bibfnamefont [1]{#1}%
\providecommand \citenamefont [1]{#1}%
\providecommand \href@noop [0]{\@secondoftwo}%
\providecommand \href [0]{\begingroup \@sanitize@url \@href}%
\providecommand \@href[1]{\@@startlink{#1}\@@href}%
\providecommand \@@href[1]{\endgroup#1\@@endlink}%
\providecommand \@sanitize@url [0]{\catcode `\\12\catcode `\$12\catcode
  `\&12\catcode `\#12\catcode `\^12\catcode `\_12\catcode `\%12\relax}%
\providecommand \@@startlink[1]{}%
\providecommand \@@endlink[0]{}%
\providecommand \url  [0]{\begingroup\@sanitize@url \@url }%
\providecommand \@url [1]{\endgroup\@href {#1}{\urlprefix }}%
\providecommand \urlprefix  [0]{URL }%
\providecommand \Eprint [0]{\href }%
\providecommand \doibase [0]{http://dx.doi.org/}%
\providecommand \selectlanguage [0]{\@gobble}%
\providecommand \bibinfo  [0]{\@secondoftwo}%
\providecommand \bibfield  [0]{\@secondoftwo}%
\providecommand \translation [1]{[#1]}%
\providecommand \BibitemOpen [0]{}%
\providecommand \bibitemStop [0]{}%
\providecommand \bibitemNoStop [0]{.\EOS\space}%
\providecommand \EOS [0]{\spacefactor3000\relax}%
\providecommand \BibitemShut  [1]{\csname bibitem#1\endcsname}%
\let\auto@bib@innerbib\@empty
\bibitem [{\citenamefont {Garbuio}\ \emph {et~al.}(2006)\citenamefont
  {Garbuio}, \citenamefont {Cascella}, \citenamefont {Reining}, \citenamefont
  {{Del Sole}},\ and\ \citenamefont {Pulci}}]{Garbuio_2006}%
  \BibitemOpen
  \bibfield  {author} {\bibinfo {author} {\bibfnamefont {V.}~\bibnamefont
  {Garbuio}}, \bibinfo {author} {\bibfnamefont {M.}~\bibnamefont {Cascella}},
  \bibinfo {author} {\bibfnamefont {L.}~\bibnamefont {Reining}}, \bibinfo
  {author} {\bibfnamefont {R.}~\bibnamefont {{Del Sole}}}, \ and\ \bibinfo
  {author} {\bibfnamefont {O.}~\bibnamefont {Pulci}},\ }\href
  {http://dx.doi.org/10.1103/PhysRevLett.97.137402} {\bibfield  {journal}
  {\bibinfo  {journal} {Phys. Rev. Lett.}\ }\textbf {\bibinfo {volume} {97}},\
  \bibinfo {pages} {137402} (\bibinfo {year} {2006})}\BibitemShut {NoStop}%
\bibitem [{\citenamefont {Ziaei}\ and\ \citenamefont
  {Bredow}(2016)}]{Ziaei_2016}%
  \BibitemOpen
  \bibfield  {author} {\bibinfo {author} {\bibfnamefont {V.}~\bibnamefont
  {Ziaei}}\ and\ \bibinfo {author} {\bibfnamefont {T.}~\bibnamefont {Bredow}},\
  }\href {\doibase 10.1063/1.4960561} {\bibfield  {journal} {\bibinfo
  {journal} {J. Chem. Phys.}\ }\textbf {\bibinfo {volume} {145}},\ \bibinfo
  {pages} {064508} (\bibinfo {year} {2016})}\BibitemShut {NoStop}%
\bibitem [{\citenamefont {Pan}\ \emph {et~al.}(2014)\citenamefont {Pan},
  \citenamefont {Wan},\ and\ \citenamefont {Galli}}]{Pan_2014}%
  \BibitemOpen
  \bibfield  {author} {\bibinfo {author} {\bibfnamefont {D.}~\bibnamefont
  {Pan}}, \bibinfo {author} {\bibfnamefont {Q.}~\bibnamefont {Wan}}, \ and\
  \bibinfo {author} {\bibfnamefont {G.}~\bibnamefont {Galli}},\ }\href
  {https://doi.org/10.1038%2Fncomms4919} {\bibfield  {journal} {\bibinfo
  {journal} {Nat. Commun.}\ }\textbf {\bibinfo {volume} {5}},\ \bibinfo {pages}
  {3919} (\bibinfo {year} {2014})}\BibitemShut {NoStop}%
\bibitem [{\citenamefont {Chan}\ \emph {et~al.}(1993)\citenamefont {Chan},
  \citenamefont {Cooper},\ and\ \citenamefont {Brion}}]{Chan_1993}%
  \BibitemOpen
  \bibfield  {author} {\bibinfo {author} {\bibfnamefont {W.}~\bibnamefont
  {Chan}}, \bibinfo {author} {\bibfnamefont {G.}~\bibnamefont {Cooper}}, \ and\
  \bibinfo {author} {\bibfnamefont {C.}~\bibnamefont {Brion}},\ }\href
  {\doibase 10.1016/0301-0104(93)85078-m} {\bibfield  {journal} {\bibinfo
  {journal} {Chem. Phys.}\ }\textbf {\bibinfo {volume} {178}},\ \bibinfo
  {pages} {387} (\bibinfo {year} {1993})}\BibitemShut {NoStop}%
\bibitem [{\citenamefont {Elles}\ \emph {et~al.}(2009)\citenamefont {Elles},
  \citenamefont {Rivera}, \citenamefont {Zhang}, \citenamefont {Pieniazek},\
  and\ \citenamefont {Bradforth}}]{Elles_2009}%
  \BibitemOpen
  \bibfield  {author} {\bibinfo {author} {\bibfnamefont {C.~G.}\ \bibnamefont
  {Elles}}, \bibinfo {author} {\bibfnamefont {C.~A.}\ \bibnamefont {Rivera}},
  \bibinfo {author} {\bibfnamefont {Y.}~\bibnamefont {Zhang}}, \bibinfo
  {author} {\bibfnamefont {P.~A.}\ \bibnamefont {Pieniazek}}, \ and\ \bibinfo
  {author} {\bibfnamefont {S.~E.}\ \bibnamefont {Bradforth}},\ }\href {\doibase
  10.1063/1.3078336} {\bibfield  {journal} {\bibinfo  {journal} {J. Chem.
  Phys.}\ }\textbf {\bibinfo {volume} {130}},\ \bibinfo {pages} {084501}
  (\bibinfo {year} {2009})}\BibitemShut {NoStop}%
\bibitem [{\citenamefont {Heller}(1974)}]{Heller_1974}%
  \BibitemOpen
  \bibfield  {author} {\bibinfo {author} {\bibfnamefont {J.~M.}\ \bibnamefont
  {Heller}},\ }\href {\doibase 10.1063/1.1681563} {\bibfield  {journal}
  {\bibinfo  {journal} {J. Chem. Phys.}\ }\textbf {\bibinfo {volume} {60}},\
  \bibinfo {pages} {3483} (\bibinfo {year} {1974})}\BibitemShut {NoStop}%
\bibitem [{\citenamefont {Hayashi}\ and\ \citenamefont
  {Hiraoka}(2015)}]{Hayashi_2015}%
  \BibitemOpen
  \bibfield  {author} {\bibinfo {author} {\bibfnamefont {H.}~\bibnamefont
  {Hayashi}}\ and\ \bibinfo {author} {\bibfnamefont {N.}~\bibnamefont
  {Hiraoka}},\ }\href {\doibase 10.1021/acs.jpcb.5b01567} {\bibfield  {journal}
  {\bibinfo  {journal} {J. Phys. Chem. B}\ }\textbf {\bibinfo {volume} {119}},\
  \bibinfo {pages} {5609} (\bibinfo {year} {2015})}\BibitemShut {NoStop}%
\bibitem [{\citenamefont {Hahn}\ \emph {et~al.}(2005)\citenamefont {Hahn},
  \citenamefont {Schmidt}, \citenamefont {Seino}, \citenamefont {Preuss},
  \citenamefont {Bechstedt},\ and\ \citenamefont {Bernholc}}]{Hahn_2005}%
  \BibitemOpen
  \bibfield  {author} {\bibinfo {author} {\bibfnamefont {P.~H.}\ \bibnamefont
  {Hahn}}, \bibinfo {author} {\bibfnamefont {W.~G.}\ \bibnamefont {Schmidt}},
  \bibinfo {author} {\bibfnamefont {K.}~\bibnamefont {Seino}}, \bibinfo
  {author} {\bibfnamefont {M.}~\bibnamefont {Preuss}}, \bibinfo {author}
  {\bibfnamefont {F.}~\bibnamefont {Bechstedt}}, \ and\ \bibinfo {author}
  {\bibfnamefont {J.}~\bibnamefont {Bernholc}},\ }\href
  {https://doi.org/10.1103%2Fphysrevlett.94.037404} {\bibfield  {journal}
  {\bibinfo  {journal} {Phys. Rev. Lett.}\ }\textbf {\bibinfo {volume} {94}},\
  \bibinfo {pages} {037404} (\bibinfo {year} {2005})}\BibitemShut {NoStop}%
\bibitem [{\citenamefont {Hermann}\ \emph {et~al.}(2008)\citenamefont
  {Hermann}, \citenamefont {Schmidt},\ and\ \citenamefont
  {Schwerdtfeger}}]{Hermann_2008}%
  \BibitemOpen
  \bibfield  {author} {\bibinfo {author} {\bibfnamefont {A.}~\bibnamefont
  {Hermann}}, \bibinfo {author} {\bibfnamefont {W.~G.}\ \bibnamefont
  {Schmidt}}, \ and\ \bibinfo {author} {\bibfnamefont {P.}~\bibnamefont
  {Schwerdtfeger}},\ }\href {https://doi.org/10.1103%2Fphysrevlett.100.207403}
  {\bibfield  {journal} {\bibinfo  {journal} {Phys. Rev. Lett.}\ }\textbf
  {\bibinfo {volume} {100}},\ \bibinfo {pages} {207403} (\bibinfo {year}
  {2008})}\BibitemShut {NoStop}%
\bibitem [{\citenamefont {Mata}\ \emph {et~al.}(2009)\citenamefont {Mata},
  \citenamefont {Stoll},\ and\ \citenamefont {Cabral}}]{Mata_2009}%
  \BibitemOpen
  \bibfield  {author} {\bibinfo {author} {\bibfnamefont {R.~A.}\ \bibnamefont
  {Mata}}, \bibinfo {author} {\bibfnamefont {H.}~\bibnamefont {Stoll}}, \ and\
  \bibinfo {author} {\bibfnamefont {B.~J.~C.}\ \bibnamefont {Cabral}},\ }\href
  {\doibase 10.1021/ct9001653} {\bibfield  {journal} {\bibinfo  {journal} {J.
  Chem. Theory Comput.}\ }\textbf {\bibinfo {volume} {5}},\ \bibinfo {pages}
  {1829} (\bibinfo {year} {2009})}\BibitemShut {NoStop}%
\bibitem [{\citenamefont {Lu}\ \emph {et~al.}(2008)\citenamefont {Lu},
  \citenamefont {Gygi},\ and\ \citenamefont {Galli}}]{Lu_2008}%
  \BibitemOpen
  \bibfield  {author} {\bibinfo {author} {\bibfnamefont {D.}~\bibnamefont
  {Lu}}, \bibinfo {author} {\bibfnamefont {F.}~\bibnamefont {Gygi}}, \ and\
  \bibinfo {author} {\bibfnamefont {G.}~\bibnamefont {Galli}},\ }\href
  {https://doi.org/10.1103%2Fphysrevlett.100.147601} {\bibfield  {journal}
  {\bibinfo  {journal} {Phys. Rev. Lett.}\ }\textbf {\bibinfo {volume} {100}},\
  \bibinfo {pages} {147601} (\bibinfo {year} {2008})}\BibitemShut {NoStop}%
\bibitem [{\citenamefont {Blase}\ \emph {et~al.}(2016)\citenamefont {Blase},
  \citenamefont {Boulanger}, \citenamefont {Bruneval}, \citenamefont
  {Fernandez-Serra},\ and\ \citenamefont {Duchemin}}]{Blase_2016}%
  \BibitemOpen
  \bibfield  {author} {\bibinfo {author} {\bibfnamefont {X.}~\bibnamefont
  {Blase}}, \bibinfo {author} {\bibfnamefont {P.}~\bibnamefont {Boulanger}},
  \bibinfo {author} {\bibfnamefont {F.}~\bibnamefont {Bruneval}}, \bibinfo
  {author} {\bibfnamefont {M.}~\bibnamefont {Fernandez-Serra}}, \ and\ \bibinfo
  {author} {\bibfnamefont {I.}~\bibnamefont {Duchemin}},\ }\href {\doibase
  10.1063/1.4940139} {\bibfield  {journal} {\bibinfo  {journal} {J. Chem.
  Phys.}\ }\textbf {\bibinfo {volume} {144}},\ \bibinfo {pages} {034109}
  (\bibinfo {year} {2016})}\BibitemShut {NoStop}%
\bibitem [{\citenamefont {Jacob}\ and\ \citenamefont
  {Neugebauer}(2014)}]{jaco2014}%
  \BibitemOpen
  \bibfield  {author} {\bibinfo {author} {\bibfnamefont {C.~R.}\ \bibnamefont
  {Jacob}}\ and\ \bibinfo {author} {\bibfnamefont {J.}~\bibnamefont
  {Neugebauer}},\ }\href {\doibase 10.1002/wcms.1175} {\bibfield  {journal}
  {\bibinfo  {journal} {WIREs: Comput. Mol. Sci.}\ }\textbf {\bibinfo {volume}
  {4}},\ \bibinfo {pages} {325} (\bibinfo {year} {2014})}\BibitemShut {NoStop}%
\bibitem [{\citenamefont {Wesolowski}\ \emph {et~al.}(2015)\citenamefont
  {Wesolowski}, \citenamefont {Shedge},\ and\ \citenamefont {Zhou}}]{weso2015}%
  \BibitemOpen
  \bibfield  {author} {\bibinfo {author} {\bibfnamefont {T.~A.}\ \bibnamefont
  {Wesolowski}}, \bibinfo {author} {\bibfnamefont {S.}~\bibnamefont {Shedge}},
  \ and\ \bibinfo {author} {\bibfnamefont {X.}~\bibnamefont {Zhou}},\ }\href
  {\doibase 10.1021/cr500502v} {\bibfield  {journal} {\bibinfo  {journal}
  {Chem. Rev.}\ }\textbf {\bibinfo {volume} {115}},\ \bibinfo {pages} {5891}
  (\bibinfo {year} {2015})}\BibitemShut {NoStop}%
\bibitem [{\citenamefont {Krishtal}\ \emph
  {et~al.}(2015{\natexlab{a}})\citenamefont {Krishtal}, \citenamefont {Sinha},
  \citenamefont {Genova},\ and\ \citenamefont {Pavanello}}]{krish2015a}%
  \BibitemOpen
  \bibfield  {author} {\bibinfo {author} {\bibfnamefont {A.}~\bibnamefont
  {Krishtal}}, \bibinfo {author} {\bibfnamefont {D.}~\bibnamefont {Sinha}},
  \bibinfo {author} {\bibfnamefont {A.}~\bibnamefont {Genova}}, \ and\ \bibinfo
  {author} {\bibfnamefont {M.}~\bibnamefont {Pavanello}},\ }\href
  {http://iopscience.iop.org/0953-8984/27/18/183202} {\bibfield  {journal}
  {\bibinfo  {journal} {J. Phys.: Condens. Matter}\ }\textbf {\bibinfo {volume}
  {27}},\ \bibinfo {pages} {183202} (\bibinfo {year}
  {2015}{\natexlab{a}})}\BibitemShut {NoStop}%
\bibitem [{\citenamefont {Casida}\ and\ \citenamefont
  {Wesolowski}(2004)}]{casi2004}%
  \BibitemOpen
  \bibfield  {author} {\bibinfo {author} {\bibfnamefont {M.~E.}\ \bibnamefont
  {Casida}}\ and\ \bibinfo {author} {\bibfnamefont {T.~A.}\ \bibnamefont
  {Wesolowski}},\ }\href@noop {} {\bibfield  {journal} {\bibinfo  {journal}
  {Int. J. Quantum Chem.}\ }\textbf {\bibinfo {volume} {96}},\ \bibinfo {pages}
  {577} (\bibinfo {year} {2004})}\BibitemShut {NoStop}%
\bibitem [{\citenamefont {Neugebauer}(2010)}]{neug2010a}%
  \BibitemOpen
  \bibfield  {author} {\bibinfo {author} {\bibfnamefont {J.}~\bibnamefont
  {Neugebauer}},\ }\href@noop {} {\bibfield  {journal} {\bibinfo  {journal}
  {Phys. Rep.}\ }\textbf {\bibinfo {volume} {489}},\ \bibinfo {pages} {1}
  (\bibinfo {year} {2010})}\BibitemShut {NoStop}%
\bibitem [{\citenamefont {Krishtal}\ \emph
  {et~al.}(2015{\natexlab{b}})\citenamefont {Krishtal}, \citenamefont
  {Ceresoli},\ and\ \citenamefont {Pavanello}}]{krish2015b}%
  \BibitemOpen
  \bibfield  {author} {\bibinfo {author} {\bibfnamefont {A.}~\bibnamefont
  {Krishtal}}, \bibinfo {author} {\bibfnamefont {D.}~\bibnamefont {Ceresoli}},
  \ and\ \bibinfo {author} {\bibfnamefont {M.}~\bibnamefont {Pavanello}},\
  }\href
  {http://scitation.aip.org/content/aip/journal/jcp/142/15/10.1063/1.4918276}
  {\bibfield  {journal} {\bibinfo  {journal} {J. Chem. Phys.}\ }\textbf
  {\bibinfo {volume} {142}},\ \bibinfo {pages} {154116} (\bibinfo {year}
  {2015}{\natexlab{b}})}\BibitemShut {NoStop}%
\bibitem [{\citenamefont {Genova}\ \emph {et~al.}(2016)\citenamefont {Genova},
  \citenamefont {Ceresoli},\ and\ \citenamefont {Pavanello}}]{Genova_2016a}%
  \BibitemOpen
  \bibfield  {author} {\bibinfo {author} {\bibfnamefont {A.}~\bibnamefont
  {Genova}}, \bibinfo {author} {\bibfnamefont {D.}~\bibnamefont {Ceresoli}}, \
  and\ \bibinfo {author} {\bibfnamefont {M.}~\bibnamefont {Pavanello}},\ }\href
  {\doibase http://dx.doi.org/10.1063/1.4953363} {\bibfield  {journal}
  {\bibinfo  {journal} {J. Chem. Phys.}\ }\textbf {\bibinfo {volume} {144}},\
  \bibinfo {pages} {234105} (\bibinfo {year} {2016})}\BibitemShut {NoStop}%
\bibitem [{\citenamefont {Genova}\ \emph {et~al.}(2017)\citenamefont {Genova},
  \citenamefont {Ceresoli}, \citenamefont {Krishtal}, \citenamefont
  {Andreussi}, \citenamefont {{DiStasio Jr.}},\ and\ \citenamefont
  {Pavanello}}]{eQE}%
  \BibitemOpen
  \bibfield  {author} {\bibinfo {author} {\bibfnamefont {A.}~\bibnamefont
  {Genova}}, \bibinfo {author} {\bibfnamefont {D.}~\bibnamefont {Ceresoli}},
  \bibinfo {author} {\bibfnamefont {A.}~\bibnamefont {Krishtal}}, \bibinfo
  {author} {\bibfnamefont {O.}~\bibnamefont {Andreussi}}, \bibinfo {author}
  {\bibfnamefont {R.}~\bibnamefont {{DiStasio Jr.}}}, \ and\ \bibinfo {author}
  {\bibfnamefont {M.}~\bibnamefont {Pavanello}},\ }\href {\doibase
  10.1002/qua.25401} {\bibfield  {journal} {\bibinfo  {journal} {Int. J.
  Quantum Chem.}\ ,\ \bibinfo {pages} {e25401}} (\bibinfo {year}
  {2017})}\BibitemShut {NoStop}%
\bibitem [{\citenamefont {Rappe}\ \emph {et~al.}(1990)\citenamefont {Rappe},
  \citenamefont {Rabe}, \citenamefont {Kaxiras},\ and\ \citenamefont
  {Joannopoulos}}]{Rappe_1990}%
  \BibitemOpen
  \bibfield  {author} {\bibinfo {author} {\bibfnamefont {A.~M.}\ \bibnamefont
  {Rappe}}, \bibinfo {author} {\bibfnamefont {K.~M.}\ \bibnamefont {Rabe}},
  \bibinfo {author} {\bibfnamefont {E.}~\bibnamefont {Kaxiras}}, \ and\
  \bibinfo {author} {\bibfnamefont {J.~D.}\ \bibnamefont {Joannopoulos}},\
  }\href {\doibase 10.1103/physrevb.41.1227} {\bibfield  {journal} {\bibinfo
  {journal} {Phys. Rev. B}\ }\textbf {\bibinfo {volume} {41}},\ \bibinfo
  {pages} {1227} (\bibinfo {year} {1990})}\BibitemShut {NoStop}%
\bibitem [{\citenamefont {Yabana}\ and\ \citenamefont
  {Bertsch}(1996)}]{Yabana_1996}%
  \BibitemOpen
  \bibfield  {author} {\bibinfo {author} {\bibfnamefont {K.}~\bibnamefont
  {Yabana}}\ and\ \bibinfo {author} {\bibfnamefont {G.~F.}\ \bibnamefont
  {Bertsch}},\ }\href {\doibase 10.1103/physrevb.54.4484} {\bibfield  {journal}
  {\bibinfo  {journal} {Phys. Rev. B}\ }\textbf {\bibinfo {volume} {54}},\
  \bibinfo {pages} {4484} (\bibinfo {year} {1996})}\BibitemShut {NoStop}%
\bibitem [{\citenamefont {Lembarki}\ and\ \citenamefont
  {Chermette}(1994)}]{lemb1994}%
  \BibitemOpen
  \bibfield  {author} {\bibinfo {author} {\bibfnamefont {A.}~\bibnamefont
  {Lembarki}}\ and\ \bibinfo {author} {\bibfnamefont {H.}~\bibnamefont
  {Chermette}},\ }\href@noop {} {\bibfield  {journal} {\bibinfo  {journal}
  {Phys. Rev. A}\ }\textbf {\bibinfo {volume} {50}},\ \bibinfo {pages} {5328}
  (\bibinfo {year} {1994})}\BibitemShut {NoStop}%
\bibitem [{\citenamefont {Bertsch}\ \emph {et~al.}(2000)\citenamefont
  {Bertsch}, \citenamefont {Iwata}, \citenamefont {Rubio},\ and\ \citenamefont
  {Yabana}}]{Bertsch_2000}%
  \BibitemOpen
  \bibfield  {author} {\bibinfo {author} {\bibfnamefont {G.~F.}\ \bibnamefont
  {Bertsch}}, \bibinfo {author} {\bibfnamefont {J.-I.}\ \bibnamefont {Iwata}},
  \bibinfo {author} {\bibfnamefont {A.}~\bibnamefont {Rubio}}, \ and\ \bibinfo
  {author} {\bibfnamefont {K.}~\bibnamefont {Yabana}},\ }\href {\doibase
  10.1103/physrevb.62.7998} {\bibfield  {journal} {\bibinfo  {journal} {Phys.
  Rev. B}\ }\textbf {\bibinfo {volume} {62}},\ \bibinfo {pages} {7998}
  (\bibinfo {year} {2000})}\BibitemShut {NoStop}%
\bibitem [{\citenamefont {Pavanello}(2013)}]{pava2013b}%
  \BibitemOpen
  \bibfield  {author} {\bibinfo {author} {\bibfnamefont {M.}~\bibnamefont
  {Pavanello}},\ }\href {\doibase http://dx.doi.org/10.1063/1.4807059}
  {\bibfield  {journal} {\bibinfo  {journal} {J. Chem. Phys.}\ }\textbf
  {\bibinfo {volume} {138}},\ \bibinfo {pages} {204118} (\bibinfo {year}
  {2013})}\BibitemShut {NoStop}%
\bibitem [{\citenamefont {Neugebauer}(2007)}]{neug2007}%
  \BibitemOpen
  \bibfield  {author} {\bibinfo {author} {\bibfnamefont {J.}~\bibnamefont
  {Neugebauer}},\ }\href@noop {} {\bibfield  {journal} {\bibinfo  {journal} {J.
  Chem. Phys.}\ }\textbf {\bibinfo {volume} {126}},\ \bibinfo {pages} {134116}
  (\bibinfo {year} {2007})}\BibitemShut {NoStop}%
\bibitem [{\citenamefont {Krishtal}\ and\ \citenamefont
  {Pavanello}(2016)}]{krish_2016}%
  \BibitemOpen
  \bibfield  {author} {\bibinfo {author} {\bibfnamefont {A.}~\bibnamefont
  {Krishtal}}\ and\ \bibinfo {author} {\bibfnamefont {M.}~\bibnamefont
  {Pavanello}},\ }\href {\doibase http://dx.doi.org/10.1063/1.4944526}
  {\bibfield  {journal} {\bibinfo  {journal} {J. Chem. Phys.}\ }\textbf
  {\bibinfo {volume} {144}},\ \bibinfo {pages} {124118} (\bibinfo {year}
  {2016})}\BibitemShut {NoStop}%
\bibitem [{\citenamefont {Onida}\ \emph {et~al.}(2002)\citenamefont {Onida},
  \citenamefont {Reining},\ and\ \citenamefont {Rubio}}]{Onida_2002}%
  \BibitemOpen
  \bibfield  {author} {\bibinfo {author} {\bibfnamefont {G.}~\bibnamefont
  {Onida}}, \bibinfo {author} {\bibfnamefont {L.}~\bibnamefont {Reining}}, \
  and\ \bibinfo {author} {\bibfnamefont {A.}~\bibnamefont {Rubio}},\ }\href
  {\doibase 10.1103/revmodphys.74.601} {\bibfield  {journal} {\bibinfo
  {journal} {Rev. Mod. Phys.}\ }\textbf {\bibinfo {volume} {74}},\ \bibinfo
  {pages} {601} (\bibinfo {year} {2002})}\BibitemShut {NoStop}%
\bibitem [{\citenamefont {Tavernelli}(2006)}]{Tavernelli_2006}%
  \BibitemOpen
  \bibfield  {author} {\bibinfo {author} {\bibfnamefont {I.}~\bibnamefont
  {Tavernelli}},\ }\href {http://dx.doi.org/10.1103/PhysRevB.73.094204}
  {\bibfield  {journal} {\bibinfo  {journal} {Phys. Rev. B}\ }\textbf {\bibinfo
  {volume} {73}} (\bibinfo {year} {2006})}\BibitemShut {NoStop}%
\bibitem [{\citenamefont {{te Velde}}\ \emph {et~al.}(2001)\citenamefont {{te
  Velde}}, \citenamefont {Bickelhaupt}, \citenamefont {Baerends}, \citenamefont
  {{van Gisbergen}}, \citenamefont {{Fonseca Guerra}}, \citenamefont
  {Snijders},\ and\ \citenamefont {Ziegler}}]{teve2001a}%
  \BibitemOpen
  \bibfield  {author} {\bibinfo {author} {\bibfnamefont {G.}~\bibnamefont {{te
  Velde}}}, \bibinfo {author} {\bibfnamefont {F.~M.}\ \bibnamefont
  {Bickelhaupt}}, \bibinfo {author} {\bibfnamefont {E.~J.}\ \bibnamefont
  {Baerends}}, \bibinfo {author} {\bibfnamefont {S.~J.~A.}\ \bibnamefont {{van
  Gisbergen}}}, \bibinfo {author} {\bibfnamefont {C.}~\bibnamefont {{Fonseca
  Guerra}}}, \bibinfo {author} {\bibfnamefont {J.~G.}\ \bibnamefont
  {Snijders}}, \ and\ \bibinfo {author} {\bibfnamefont {T.}~\bibnamefont
  {Ziegler}},\ }\href@noop {} {\bibfield  {journal} {\bibinfo  {journal} {J.
  Comput. Chem.}\ }\textbf {\bibinfo {volume} {22}},\ \bibinfo {pages} {931}
  (\bibinfo {year} {2001})}\BibitemShut {NoStop}%
\bibitem [{\citenamefont {DiStasio}\ \emph {et~al.}(2014)\citenamefont
  {DiStasio}, \citenamefont {Santra}, \citenamefont {Li}, \citenamefont {Wu},\
  and\ \citenamefont {Car}}]{DiStasio_2014b}%
  \BibitemOpen
  \bibfield  {author} {\bibinfo {author} {\bibfnamefont {R.~A.}\ \bibnamefont
  {DiStasio}}, \bibinfo {author} {\bibfnamefont {B.}~\bibnamefont {Santra}},
  \bibinfo {author} {\bibfnamefont {Z.}~\bibnamefont {Li}}, \bibinfo {author}
  {\bibfnamefont {X.}~\bibnamefont {Wu}}, \ and\ \bibinfo {author}
  {\bibfnamefont {R.}~\bibnamefont {Car}},\ }\href {\doibase 10.1063/1.4893377}
  {\bibfield  {journal} {\bibinfo  {journal} {J. Chem. Phys.}\ }\textbf
  {\bibinfo {volume} {141}},\ \bibinfo {pages} {084502} (\bibinfo {year}
  {2014})}\BibitemShut {NoStop}%
\end{thebibliography}
%

\end{document}